\documentclass[11pt,twoside]{article}
\usepackage{./asp2014}

\aspSuppressVolSlug
\resetcounters

\begin{document}

\title{Spectroscopic diagnostics for circumstellar disks of B[e] supergiants}
\author{M. Kraus,$^{1,2}$ 
\affil{$^1$Astronomick\'{y} \'{u}stav AV\,\v{C}R, v.v.i., Ond\v{r}ejov, Czech Republic; \email{michaela.kraus@asu.cas.cz}\\
$^2$Tartu Observatory, T\~oravere, Estonia}}

\paperauthor{Michaela~Kraus}{michaela.kraus@asu.cas.cz}{}{Astronomick\'y \'ustav, Akademie v\v{e}d \v{C}esk\'e republiky, v.v.i.}{}{Ond\v{r}ejov}{}{25165}{Czech Republic}


\begin{abstract}
B[e] supergiants (B[e]SGs) are emission-line objects, presumably in a 
short-lived phase in the post-main sequence evolution of massive stars. 
Their intense infrared excess emission indicates large amounts of warm 
circumstellar dust, and the stars were longtime assumed to possess an 
aspherical wind consisting of a classical line-driven wind in polar direction 
and a dense, slow equatorial wind dubbed outflowing disk. The general 
properties obtained for these disks are in line with this scenario, although 
current theories have considerable difficulties reproducing the
observed quantities. Therefore, more sophisticated observational constraints 
are needed. These follow from combined optical and infrared spectroscopic 
studies, which delivered the surprising result that the circumstellar material 
of B[e]SGs is concentrated in multiple rings revolving the stars on stable 
Keplerian orbits. Such a scenario requires new ideas for the formation 
mechanism, in which pulsations might play an important role.
\end{abstract}

\section{Introduction}

B[e] supergiants (B[e]SGs) form a sub-group of stars displaying the B[e] 
phenomenon. This phenomenon was originally discovered by 
\citet{1970ApJ...161L.105G}, who detected infrared (IR) excess emission due to
hot or warm dust in a sample of B-type emission-line stars known to display 
low-excitation forbidden and permitted emission lines (predominantly from 
Fe\,{\sc ii}). This IR excess emission was confirmed by follow-up surveys 
\citep{1972ApL....10...83A, 1973MNRAS.161..145A, 1974MNRAS.168....1A, 
1974MNRAS.167..337A, 1975MNRAS.170..579A}. Inspection of these objects 
revealed that the sample consists of stars in different evolutionary phases,
including pre-main sequence stars, compact planetary nebulae, supergiants, and 
symbiotic binaries.

Based in the classification criteria defined by \citet{Lamers1998}, a 
star with the B[e] phenomenon is assigned a B[e]SG status, when it fulfills 
additional criteria. These are a supergiant luminosity ($\log L/L_{\sun} \ge 
4.0$), small photometric variability, chemically processed material indicating 
an evolved nature, indications of mass loss in the optical spectra, and a 
hybrid spectrum consisting of narrow low-excitation emission lines of 
low-ionized metals (e.g., Fe\,{\sc ii}, [Fe\,{\sc ii}], [O\,{\sc i}]) with 
simultaneous broad absorption features of high-excitation lines. 

The number of confirmed B[e]SGs is small. In the Galaxy, it is difficult to
assign a star a firm supergiant status due to the often uncertain distances,
hence luminosities. Previously classified stars might also suddenly display
a variable character, which then requires re-classification of the object
\citep[e.g.,][]{2016MNRAS.456.1424A}. It might thus be advisable to speak of
candidates only, and there are currently about 16 such Galactic B[e]SG
candidates known \citep[for a list see][plus the stars HD\,62623 and
AS\,381]{2009A&A...494..253K}.

The Magellanic Clouds host the best-known B[e]SG
sample. Here, the proper luminosity assignment is not an issue. We know of 15
confirmed members \citep{2006ASPC..355..135Z}, of which eleven reside in the
Large Magellanic Cloud (LMC) and four in the Small
Magellanic Cloud (SMC). In addition, a few new B[e]SG candidates were
discovered: three in the LMC \citep[see][]{2012A&A...542A..50D,
2014A&A...568A..28L} and four in the SMC \citep[see][]{2007ApJ...670.1331W,
2012ApJ...759...10G}. Moreover, \citet{2015MNRAS.454.1468K} recently reported
on a new sample of B[e] candidates in the LMC. Whether also supergiants are
included in this sample still needs to be investigated.

The identification of B[e]SGs in other galaxies is difficult, because these 
objects are generally rather faint due to the large distances. Nevertheless, 
optical spectroscopic surveys of evolved massive stars in Local Group Galaxies 
hinted at a number of possible candidates \citep[e.g.,][]{2007AJ....134.2474M, 
2012A&A...541A.146C}. \citet{2014ApJ...780L..10K} performed follow-up 
IR observations of a few, selected candidates. Based on the IR appearance of the
stars, they discovered the first two B[e]SGs in the Andromeda galaxy. This 
demonstrates that a large number of these peculiar objects might still be 
identified in the Local Group with a more systematic search.

\section{General properties of B[e] supergiant stars' disks}

The hybrid spectrum of B[e]SGs was discovered by \citet{1985A&A...143..421Z}. 
These authors proposed that B[e]SGs might have a hybrid wind, consisting of a 
classical (i.e., low-density, fast) line-driven wind in polar direction 
together with an equatorial high-density, slow outflow. For these hybrid winds, 
a density contrast of 100 to 1000 between the equatorial and the polar wind 
components was found \citep{1989A&A...220..206Z}, and the equatorial wind 
was henceforth dubbed ``outflowing disk''.

The evidences for dense and cool disk-like structures around B[e]SGs are 
mani\-fold. Their spectral energy distributions display strong IR emission in 
excess of a pure free-free contribution from a stellar wind or ionized gas disk
\citep[see, e.g.,][]{1986A&A...163..119Z, 2009AJ....138.1003B, 
2010AJ....140..416B}. It is thus attributed to large amounts of circumstellar 
dust. As the IR emission peaks in the near-IR region, this dust must be rather 
hot ($\sim 1000$\,K). The spatial distribution of the dust cannot be 
spherically symmetric because \citet{1992ApJ...398..286M} discovered intrinsic 
optical linear polarization in all observed objects. The good correlation 
between the degree of polarization and the intensity of the dust emission he 
assigned to dusty disks seen under different inclination angles. Other studies
confirmed the often high degree of intrinsic polarization, and detailed 
investigations suggest that it is caused by the combined effect of 
electron scattering plus scattering by dust in a disk-like structure
\citep[see,][and Seriacopi et al., this volume]{1989A&A...214..274Z}.

A dense disk is also considered as the ideal environment for the hot molecules
that were discovered in B[e]SGs. The most prominent species is hereby carbon 
monoxide (CO). Its first-overtone band emission arises in the near-IR and was
detected from many B[e]SGs \citep{1988ApJ...324.1071M, 1988ApJ...334..639M, 
1989A&A...223..237M, 1996ApJ...470..597M, 2012MNRAS.426L..56O, 
2013A&A...558A..17O}. The excitation of these bands requires gas temperatures 
in excess of 2000\,K, implying that CO band emission originates from 
circumstellar regions closer to the star than the dust.  With a dissociation 
temperature of $\sim 5000$\,K, CO is the most stable molecule in the universe, 
and no other molecule can exist closer to a hot, luminous star than CO. Hence, 
the detection of CO band emission from an object is typically regarded as the 
ultimate tracer for the inner rim of the circumstellar molecular gas. As the
strong  ultraviolet radiation field of hot stars destroys CO, 
the molecules are most likely located within a dense disk or ring, capable for
self-shielding of the gas.

Other molecules detected from a few B[e]SGs are silicon oxide (SiO), whose 
first-overtone bands also arise in the near-IR region 
\citep{2015ApJ...800L..20K}, and titanium oxide (TiO) with band emission from
electronic transitions emitting at optical wavelengths 
\citep{1989A&A...220..206Z, 2012MNRAS.427L..80T, 2016A&A...593..112K}.
However, so far not many objects have really been searched for emission from
molecules other than CO, and one might expect to find emission from (hot) 
molecules in many more B[e]SGs.

The development of optical interferometry facilities, which operate
in the near- and mid-IR wavelength regions, provided a great step forward
in B[e]SG star research (see Meilland, this volume). The capability of these
facilities to spatially resolve circumstellar environments down to small scales 
boosted the studies of B[e]SG star's disks, at least for those Galactic objects 
that were close and IR bright enough to be observed interferometrically.
With this powerful tool, the gaseous, dusty disks of several objects could
be resolved, resulting in precise measurements of the disk sizes, the 
distances of the dust from the stars, and the disk inclination angles
\citep[see][]{2007A&A...464...81D, 2011A&A...525A..22D, 2011A&A...526A.107M,
2012A&A...548A..72C, 2012A&A...545L..10W, 2012A&A...538A...6W,
2012A&A...543A..77W}. 
  
All these evidences listed above imply that B[e]SGs are surrounded by 
disk-like structures, which must be geometrically thick to provide an ideal 
dense and cool enough environment for molecule and dust condensation. A proof, 
that the material forming these disks was indeed ejected from the star during
its post-main sequence evolution, was provided by the discovery of enhanced
$^{13}$C, detected via strong $^{13}$CO band emission 
\citep{2010MNRAS.408L...6L, 2013A&A...549A..28K, 2013A&A...558A..17O}. As 
stellar evolution models depict \citep[see, e.g.][]{2012A&A...537A.146E}, the 
enrichment of the stellar surface (and thus of the released circumstellar 
material) with this isotope, with respect to the main isotope $^{12}$C, is a 
process that starts already during the main-sequence evolution of massive stars 
and continues gradually with the age of the object. Hence, detection of 
an enrichment in $^{13}$CO in stars with circumstellar disks provides the ideal
tool to estimate the evolutionary state, but also to discriminate between a 
pre-main sequence and an evolved nature of the object
\citep[e.g.,][]{2009A&A...494..253K, 2015AJ....149...13M}.

\section{Disk formation mechanisms}

As B[e]SGs are hot, luminous objects, how can cool and dense disks form and 
survive in such a harsh environment? Several mechanisms were proposed in the
literature, which are briefly outlined in the following. 

Most popular is certainly the binary scenario, in which a disk-like structure 
can form from the material ejected during phases of strong interaction, up to 
full merger. Up to now, a companion was identified in only six B[e]SGs.
Four of them are Galactic objects (MWC\,300, HD\,62623, 
HD\,327083, and GG\,Car), and in all of them the disk is circumbinary
\citep{2012A&A...545L..10W, 2011A&A...526A.107M, 2012A&A...538A...6W,
2013A&A...549A..28K}. The remaining two objects (LHA\,115-S\,6 and 
LHA\,115-S\,18) reside in the SMC, and these objects show clear differences to 
the Galactic binary sources. LHA\,115-S\,6 was suggested to be a 
post-merger candidate in an original triple system \citep{1998ASSL..233..235L, 
2006ASPC..355..259P}, while LHA\,115-S\,18 was identified as the optical
counterpart of a high-mass X-ray source \citep{2013A&A...560A..10C, 
2014MNRAS.438.2005M}. The latter object is particularly interesting, as it 
shows strong photometric and spectroscopic variability and displays 
time-variable Raman scattered emission \citep{2012MNRAS.427L..80T}, which is 
typically only seen in symbiotic systems \citep[e.g.,][]{2016MNRAS.456.2558L}. 
The features and variabilities of the
SMC B[e]SG binary candidates clearly separate these objects from the remaining 
B[e]SG sample in the Magellanic Clouds. Together with the fact that no binary 
was identified in the LMC B[e]SG sample so far, we might conclude that binary
interaction or merger might be a reasonable scenario for a minority of
objects, but it is most likely not the universal disk-formation mechanism.

For single B[e]SGs, \citet{1993ApJ...409..429B} proposed a model in which the
winds emanating from the two hemispheres of a rapidly rotating star were 
predicted to collide in the equatorial plane, forming a compressed, outflowing 
disk. However, these authors did not include the non-radial forces that arise 
from the flattening of a rapidly rotating star, which prevent the formation
of a disk \citep{1996ApJ...472L.115O}. 
Alternatively, critical stellar rotation may form 
an outflowing disk via the magneto-rotational instability mechanism 
\citep[][and this volume]{2015A&A...573A..20K, 2014A&A...569A..23K} 
This model works well for classical Be stars, which
are known to rotate (close to) critical.

To test whether the material in the equatorial plane can really recombine in 
the vicinity of a hot, luminous supergiant, ionization structure calculations 
have been performed in hybrid-wind models. Based on a heuristic approach for 
the latitudinal density distribution, recombination in the equatorial plane 
occurred, but only in models with (unrealistically?) high values for the 
equatorial mass-loss rates \citep{2003A&A...405..165K, 2008A&A...478..543Z}. 
Inclusion of (rapid) stellar rotation into the calculations revealed that due 
to the effects of gravitational darkening, the equatorial wind regions are depleted, resulting in hydrogen-neutral equatorial zones, which are less dense
than the polar regions \citep{2006A&A...456..151K}. This is the opposite of 
what is observed.
  
A way out of this dilemma is provided by the rotationally induced bi-stability 
mechanism \citep{2000A&A...359..695P}, which 
utilizes the fact that ions recombine at a certain latitude due to the latitude 
dependence of the temperature in the wind of a (rapidly) rotating star. 
Of particular importance is the recombination from Fe\,{\sc iv} to 
Fe\,{\sc iii} at a temperature of $\sim 25\,000$\,K, because Fe\,{\sc iii} 
possesses many more transitions suitable to drive a dense wind in the 
cooler, i.e. equatorial regions. The drawback of this mechanism is, however, 
that the density contrast between equatorial and polar wind that can be 
achieved is a factor 10--100 below the observed values. The situation can be 
improved by combining the bi-stability mechanism with the slow-wind solution 
discovered by \citet[][see also Cur\'{e}, this volume]{2004ApJ...614..929C}. 
In this model, proper density enhancements in the equatorial direction could be 
achieved \citep{2005A&A...437..929C}, however, the resulting wind velocities in 
the equatorial plane (200--300\,km\,s$^{-1}$) were too high compared to the 
observed values, which are on the order of 10--30\,km\,s$^{-1}$ and can display
a slow-down with increasing distance from the star 
\citep[e.g.,][]{2010A&A...517A..30K}. 

To summarize, all single-star models presented have significant drawbacks and
have as major prerequisite that B[e]SGs  should be rotating at a substantial 
fraction of their critical velocity.

\section{On the rotation velocities of B[e]SGs}

The rotation velocities of stars, projected to the line-of-sight, are typically 
extracted from the profiles of their photospheric absorption lines. As most 
B[e]SGs are deeply embedded in their dense winds, their typically pure 
emission-line spectra hamper the proper establishment of their rotation 
speeds. Only in four Magellanic Cloud
stars one or more photospheric lines were detected, that were used to estimate 
the rotation velocity, projected to the line-of-sight. These are 
LHA\,115-S\,65 \citep{2000ASPC..214...26Z, 2010A&A...517A..30K} and
LHA\,115-S\,23 \citep{2008A&A...487..697K} in the SMC, and LHA\,120-S\,73
\citep{2006ASPC..355..135Z} and LHA\,120-S\,93 \citep{1995A&A...302..409G} in
the LMC. The inclination angle is often unknown or uncertain, so that only
lower limits of their real rotation velocities are obtained. For 
LHA\,120-S\,93, a value of $\sim 40\%$ of its critical rotation 
speed was estimated, whereas for the other objects a value on the order of 
$75\%$ was proposed. Although these high values are tempting and have been used 
in the past as base for the theoretical models presented above, their 
reliability has recently been questioned. 

For the LMC star LHA\,120-S\,73, \citet{2006ASPC..355..135Z} determined a 
value of $\varv\sin i = 50$\,km\,s$^{-1}$ from the photospheric He{\sc i} 
$\lambda$ 5876. This line was also seen in the spectra
taken between 1999 and 2015. However, \citet{2016A&A...593..112K} have shown
that its profile was clearly asymmetric and time-variable. Moreover, comparison 
with an artificial rotationally-broadened line profile revealed that the core 
of the observed line is in most cases much narrower than the theoretical one, 
meaning that the value of $\varv\sin i = 50$\,km\,s$^{-1}$, and hence of $75\%$ 
critical, are significantly overestimated. The He{\sc i} $\lambda$ 5876 line 
is thus not suitable to reliably determine $\varv\sin i$ in this object. In 
light of these findings, a revision of the $\varv\sin i$ values of the 
other three objects appears inevitable.

\section{Disc structure and kinematics}

There could be two reasons why no common scenario was identified so far for
the disk formation in B[e]SGs. Either the mechanism is different in each 
object, or we did not find the proper one yet. To unveil possible disk 
formation mechanisms, suitable observational constraints regarding the 
structure and kinematics of the material surrounding B[e]SGs are
indispensable. Spectroscopy provides the ideal tool to search for features 
that trace the disk from its inner rim to far distances.

\subsection{Dust}

Let's start with the outer disk regions, dominated by dust. 
IR photometric observations over several decades displayed no evidence for 
variability. If the dust particles condense from an outflow, they should be 
formed continuously. However, as model computations have shown, the observed 
intense IR excess emission of B[e]SGs cannot be reproduced by such a scenario 
\citep{2003A&A...398..631P}. This means that not enough dust can form in situ 
in an outflowing disk. Instead, it seems more likely that the dust around 
B[e]SGs has accumulated with time. Support for such a scenario comes from 
observations with the {\it Spitzer Space Telescope}. The mid-IR spectra of the 
nine observed B[e]SGs in the Magellanic Clouds displayed emission features from 
crystalline silicates \citep{Kastner2010}. Their formation requires
substantial grain processing in a stable, long-lived environment. Some of the 
objects showed additional emission
from polycyclic aromatic hydrocarbons (PAHs). This dual-dust chemistry, i.e., 
the co-existence of silicates with carbon-based grains, is a further strong
indicator for a stable dusty environment, in which non-equilibrium chemical 
processes had sufficient time to occur.

\subsection{Molecules and chemistry}

Closer to the star, where the temperature exceeds the dust evaporation 
temperature of $\sim 1500$\,K, resides the molecular gas region. As mentioned 
earlier, molecular 
emission, in particular from CO, was detected in many B[e]SGs. The 
intensity of the CO first-overtone bands, which arise in the near-IR regime 
red-wards of 2.293\,$\mu$m, is extremely sensitive to the density and 
temperature of the CO gas \citep{2009A&A...494..253K}. As CO is furthermore the
most stable molecule, the observed CO band emission originates 
from the densest and hottest molecular region, which marks at the same time the
transition from atomic to molecular gas and hence the inner rim 
of the hot molecular disk. 

Using the SINFONI spectrograph at ESO, a $K$-band near-IR survey was performed 
to determine the physical parameters of the CO band emitting regions in a
sample of Galactic and Magellanic Cloud B[e]SGs \citep{2010MNRAS.408L...6L, 
2013A&A...558A..17O}. This survey delivered two surprising results. The CO gas 
in each object had a different temperature, spreading from 1900\,K to 3200\,K, 
and these values are far below the CO dissociation temperature of 5000\,K.
If the gas was distributed in a homogeneous disk ranging from the stellar
surface to far distances, the transition from the atomic to the molecular
region should take place at a temperature close to the CO dissociation value. 
The rather cool CO temperature values found in all objects imply that no hotter 
molecular gas exists closer to the star. Consequently the disks of B[e]SGs
cannot be continuous structures. 

The resolution of the SINFONI spectrograph is low ($R \sim 4500$). To resolve
the kinematics stored in the shape of the first CO band head, high-resolution
observations are required. So far, eight Galactic B[e]SGs were observed with
the CRIRES spectrograph at ESO, and another four Magellanic Cloud B[e]SGs with
the Phoenix spectrograph at GEMINI-South. Both instruments provide a resolution
of $R \sim 50\,000$. In all objects, the CO band head displayed a blue-shifted
``shoulder'' and a red-shifted maximum. Detailed modeling of the band head 
structures revealed that the individual lines forming the band head display
a double-peaked profile, which corresponds to just one single line-of-sight
velocity \citep[see][]{2012BAAA...55..123M, 2012A&A...543A..77W, 
2012A&A...548A..72C, 2013A&A...549A..28K, 2016A&A...593..112K}. Such a profile 
can originate either from a ring expanding with constant velocity, or 
from a very narrow rotating ring. 

To discriminate whether the velocity seen in the CO bands is due to expansion 
or rotation, information from complimentary tracers of the kinematics is needed.
Most valuable would be other molecules, which form at lower temperatures than 
CO and hence at larger distances from the star. Considering that B[e]SGs are 
massive stars, which preserve an oxygen-rich surface composition throughout 
their lifetime, a very promising molecule that will form in such an oxygen-rich 
environment is SiO. This molecule has, together with TiO, the second highest 
dissociation temperature and can hence also be regarded as very robust. 
Moreover, Si has an abundance that is more than two orders of magnitude higher 
than that of Ti, guaranteeing that a sufficiently large number of SiO will be 
formed to produce observable amounts of emission.

The structure of the energy levels in SiO is very similar to CO, so that
in a hot enough environment ($T > 1500$\,K) the rotation-vibration bands are
excited. The first-overtone SiO bands arise in the $L$-band, red-wards of 
4.004\,$\mu$m. \citet{2015ApJ...800L..20K} selected four Galactic B[e]SGs
with confirmed CO band emission, and observed these stars using 
CRIRES. All four objects displayed kinematically 
broadened emission of the first SiO band head. As the velocity projected to the
line-of-sight was smaller than what was derived from CO, the logical conclusion 
was that the molecular gas is rotating around the central stars, most likely 
on Keplerian orbits.  

The successful detection of theoretically predicted SiO band emission from the
environments of B[e]SGs encourages to search for further molecular species, 
which would bridge the gap to the dust condensation zone, and would help to 
greatly improve our knowledge on the physical properties and the kinematics of 
the molecular regions.

\subsection{Optical forbidden lines}

We turn now to the atomic gas, which is located between the stellar surface
and the molecular CO ring. To study the gas kinematics, optically thin lines 
are required. This predestines forbidden lines, because transitions from 
various elements and ionization states are excellent tracers for different 
density and temperature regimes \citep[e.g.,][]{2005A&A...441..289K}.

Besides the numerous [Fe{\sc ii}] lines in the spectra of B[e]SGs, which trace
low-density environments, the [O{\sc i}] lines, which are inherent in all 
B[e]SGs, have proven to be extremely valuable. The optical spectral range hosts 
three of them: [O{\sc i}] $\lambda\lambda$ 5577, 6300, 6364. However, the 
[O{\sc i}] $\lambda$ 5577 is often very weak or even absent. These lines
had been ignored until \citet{2007A&A...463..627K} realized that they arise
from regions in which the particle density can be high, as long as the 
electron density remains low. Such conditions prevail in regions in which 
hydrogen is predominantly neutral. This is achieved for temperatures below
$\sim 10\,000$\,K.
The [O{\sc i}] lines in high-resolution optical spectra displayed 
double-peaked profiles in B[e]SGs with edge-on or intermediate orientation of 
their disks, while they displayed no kinematical broadening in pole-on oriented 
objects \citep{2010A&A...517A..30K, 2012MNRAS.423..284A, 2012ASPC..464...67M}. 
This implies that the [O{\sc i}] line forming region lies within the dense, 
neutral (in hydrogen) regions of the gaseous disks. Moreover, the profile of 
the [O{\sc i}] $\lambda$ 5577 line is typically wider than the one of, e.g., 
[O{\sc i}] $\lambda$ 6300. As the [O{\sc i}] $\lambda$ 5577 line 
emerges from a higher energy level whose collisional population requires an 
environment with higher electron density, the line forming 
regions of [O{\sc i}] $\lambda$ 5577 and [O{\sc i}] $\lambda$ 6300 are 
physically distinct. These findings agree with the idea that the [O{\sc i}] 
lines form in a Keplerian rotating gas disk, in which the [O{\sc i}] $\lambda$ 
5577 line forms at distances closer to the star where both the density and 
rotation speed are larger.

In addition to the [O{\sc i}] lines, \citet{2012MNRAS.423..284A} recently
identified the two [Ca\,{\sc ii}] $\lambda\lambda$ 7291, 7324 lines in the 
spectra of B[e]SGs. These lines form a valuable complementary set of tracers
for the properties of the gas disk. Alike [O{\sc i}], the [Ca\,{\sc ii}] lines 
display double-peaked profiles, indicating rotational broadening, and in all
studied objects the width of the [Ca\,{\sc ii}] line profiles was comparable 
to or even broader than the one of the [O{\sc i}] $\lambda$ 5577 lines. This 
implies that the [Ca\,{\sc ii}] lines form at similar distances or even 
closer to the star than [O{\sc i}] $\lambda$ 5577 \citep{2012MNRAS.423..284A, 
2016MNRAS.456.1424A}. This is in line with the higher critical electron density
for [Ca\,{\sc ii}].

Recent detailed investigation of the [O{\sc i}] and [Ca\,{\sc ii}] sets of 
forbidden lines revealed that their profiles cannot be modeled under the 
assumption of line formation in a single ring as it was the case for CO. 
Instead, to properly reproduce 
the shapes of the profiles the emission must originate from at least two 
detached and physically distinct rings \citep[e.g.,][and this 
volume]{2016A&A...593..112K, 2015EAS....71..229M}. This result appears
surprising at first glance, but it is in line with the findings from the 
CO molecular emission, which also arises from a detached ring. 
Consequently, B[e]SGs cannot have disks formed from a continuous
equatorially outflowing wind. Instead, all tracers hint towards 
multiple material concentrations in form of rings of 
atomic and molecular gas, revolving the stars on stable orbits.

\section{CO variability and inhomogeneities}

Despite their importance, observations in the (near) IR are sparse. While most
B[e]SGs were observed in the $K$-band during the past 30 
years at least once \citep{1988ApJ...324.1071M, 1988ApJ...334..639M, 
1989A&A...223..237M, 1996ApJ...470..597M, 2013A&A...558A..17O}, most spectra 
were taken with different (mainly low) resolution. They hamper precise 
comparison of the strength and shape of the band heads and are inadequate for 
kinematical studies. Follow-up High-resolution observations exist only for a 
few objects, but even with this small set of data notable variabilities 
in the CO band emission could be identified in six objects.

The CO variabilities seen in two objects were exceptional. The first one is the 
Galactic object CI\,Cam, which experienced a spectacular outburst in 1998. And
about one month later, CO band emission was discovered 
\citep{1999A&A...348..888C}, which disappeared again with the expansion and 
dilution of the ejected material \citep{2014MNRAS.443..947L}. 
And the SMC star LHA\,115-S\,65 suddenly displayed intense CO band emission in 
October 2011, while all previous observations resulted in non-detections
\citep{2012MNRAS.426L..56O}. Whether the appearance of CO band emission in 
LHA\,115-S\,65 was also caused by an outburst event, and whether the 
emission disappeared again, is unknown. 
 
\articlefigure[width=1.0\textwidth]{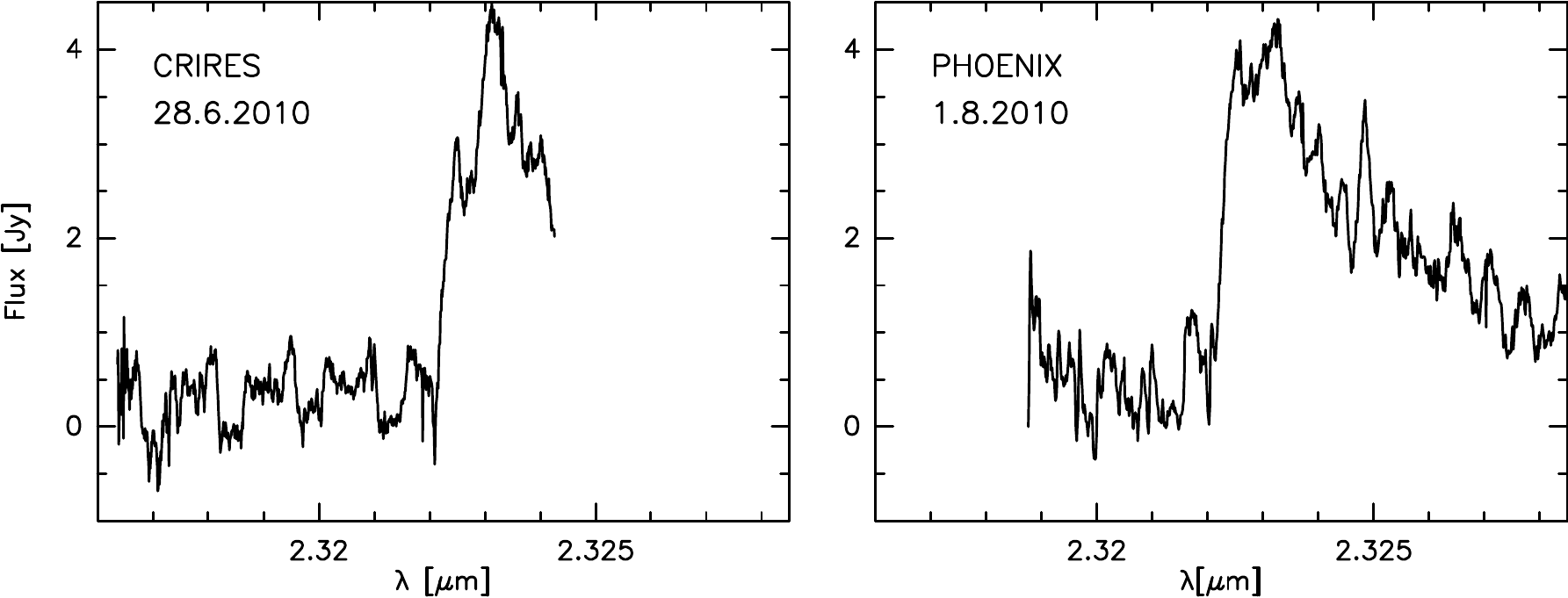}{fig}{Variation in the CO band emission observed in HD\,327083.}

The variabilities seen in the other objects are less dramatic, but still of 
high importance. The LMC star LHA\,120-S\,35 displayed a change in CO intensity 
by factor of two within two years (Torres et al., in preparation). 
LHA\,120-S\,73 exhibited a decrease in CO intensity between 2004 and 2009, 
while in 2010 the intensity had partly recovered \citep{2016A&A...593..112K}. 
And the Galactic B[e]SG HD\,327083 displayed variations in its CO band heads 
within a period of one month \citep[][see Fig.\,\ref{fig}]{2013msao.confE.160K}.
In the latter two objects the observed varibilities can be explained with 
density inhomogeneities (i.e. clumps) within the molecular gas ring. This 
scenario is supported by the spatially resolved integral-field unit $K$-band 
images of the Galactic object MWC\,137, which revealed multiple, clumpy 
molecular ring structures (Kraus et al., this volume).

\section{Conclusions and future perspectives}

Although B[e]SGs have been investigated for several decades, they are still 
puzzling. Thorough analysis of their optical and IR spectroscopic appearance 
has led to the conclusion that these objects have multiple, often clumpy ring 
structures revolving the star on presumably stable orbits, meaning that the
original suggestion of an outflowing disk needs to be discarded. 
Another important conclusion refers to the rapid (close to critical) rotation
speeds of B[e]SGs. Based on the recent findings of \citet{2016A&A...593..112K},
the values found in the literature might be significantly overestimated.
A revision of all presently known $\varv\sin i$ values is thus imperative.

The next big question that hence needs to be addressed concerns the origin of 
the circumstellar rings. While in some objects they might result from
binary interaction, most B[e]SGs are not known (yet) to have a close companion. 
For these stars an alternative formation mechanism is required. A possible 
approach might involve stellar pulsations, which were recently discovered in 
classical blue supergiants and $\alpha$\,Cygni variables 
\citep{2013MNRAS.433.1246S, 2015A&A...581A..75K}. Whether specific pulsation 
modes might facilitate periodically triggered (equatorial) mass-loss is 
certainly worth being investigated in more detail.

\acknowledgements I wish to thank the organizers for the invitation to present
this review. The shown spectra were obtained under ESO program ID 385.D-0613(A)
and GEMINI program ID GS-2010A-Q-41. This work was supported by GA\,\v{C}R 
(14-21373S) and ETAg (IUT40-1). The Astronomical Institute Ond\v{r}ejov is 
supported by the project RVO:67985815. 

\bibliography{Kraus_bep2016_review}  

\begin{thebibliography}{}
\expandafter\ifx\csname natexlab\endcsname\relax\def\natexlab#1{#1}\fi
\expandafter\ifx\csname url\endcsname\relax
  \def\url#1{\texttt{#1}}\fi
\expandafter\ifx\csname urlprefix\endcsname\relax\def\urlprefix{URL }\fi
\providecommand{\eprint}[2][]{\url{#2}}

\bibitem[{{Allen}(1973)}]{1973MNRAS.161..145A}
{Allen}, D.~A. 1973, \mnras, 161, 145

\bibitem[{{Allen}(1974)}]{1974MNRAS.168....1A}
--- 1974, \mnras, 168, 1

\bibitem[{{Allen} \& {Glass}(1974)}]{1974MNRAS.167..337A}
{Allen}, D.~A., \& {Glass}, I.~S. 1974, \mnras, 167, 337

\bibitem[{{Allen} \& {Glass}(1975)}]{1975MNRAS.170..579A}
--- 1975, \mnras, 170, 579

\bibitem[{{Allen} \& {Swings}(1972)}]{1972ApL....10...83A}
{Allen}, D.~A., \& {Swings}, J.~P. 1972, Astrophys. Lett., 10, 83

\bibitem[{{Aret} et~al.(2012){Aret}, {Kraus}, {Muratore}, \& {Borges
  Fernandes}}]{2012MNRAS.423..284A}
{Aret}, A., {Kraus}, M., {Muratore}, M.~F., \& {Borges Fernandes}, M. 2012,
  \mnras, 423, 284

\bibitem[{{Aret} et~al.(2016){Aret}, {Kraus}, \& {{\v
  S}lechta}}]{2016MNRAS.456.1424A}
{Aret}, A., {Kraus}, M., \& {{\v S}lechta}, M. 2016, \mnras, 456, 1424

\bibitem[{{Bjorkman} \& {Cassinelli}(1993)}]{1993ApJ...409..429B}
{Bjorkman}, J.~E., \& {Cassinelli}, J.~P. 1993, \apj, 409, 429

\bibitem[{{Bonanos} et~al.(2010){Bonanos}, {Lennon}, {K{\"o}hlinger}
  et~al.}]{2010AJ....140..416B}
{Bonanos}, A.~Z., {Lennon}, D.~J., {K{\"o}hlinger}, F., et~al. 2010, \aj, 140,
  416

\bibitem[{{Bonanos} et~al.(2009){Bonanos}, {Massa}, {Sewilo}
  et~al.}]{2009AJ....138.1003B}
{Bonanos}, A.~Z., {Massa}, D.~L., {Sewilo}, M., et~al. 2009, \aj, 138, 1003

\bibitem[{{Cidale} et~al.(2012){Cidale}, {Borges Fernandes}, \&
  {Andruchow}}]{2012A&A...548A..72C}
{Cidale}, L.~S., {Borges Fernandes}, M., \& {Andruchow}, I. e.~a. 2012, \aap,
  548, A72

\bibitem[{{Clark} et~al.(2013){Clark}, {Bartlett}, {Coe}
  et~al.}]{2013A&A...560A..10C}
{Clark}, J.~S., {Bartlett}, E.~S., {Coe}, M.~J., et~al. 2013, \aap, 560, A10

\bibitem[{{Clark} et~al.(2012){Clark}, {Castro}, {Garcia}
  et~al.}]{2012A&A...541A.146C}
{Clark}, J.~S., {Castro}, N., {Garcia}, M., et~al. 2012, \aap, 541, A146

\bibitem[{{Clark} et~al.(1999){Clark}, {Steele}, {Fender}, \&
  {Coe}}]{1999A&A...348..888C}
{Clark}, J.~S., {Steele}, I.~A., {Fender}, R.~P., \& {Coe}, M.~J. 1999, \aap,
  348, 888

\bibitem[{{Cur{\'e}}(2004)}]{2004ApJ...614..929C}
{Cur{\'e}}, M. 2004, \apj, 614, 929

\bibitem[{{Cur{\'e}} et~al.(2005){Cur{\'e}}, {Rial}, \&
  {Cidale}}]{2005A&A...437..929C}
{Cur{\'e}}, M., {Rial}, D.~F., \& {Cidale}, L. 2005, \aap, 437, 929

\bibitem[{{Domiciano de Souza} et~al.(2011){Domiciano de Souza}, {Bendjoya}, \&
  {Niccolini}}]{2011A&A...525A..22D}
{Domiciano de Souza}, A., {Bendjoya}, P., \& {Niccolini}, G. e.~a. 2011, \aap,
  525, A22

\bibitem[{{Domiciano de Souza} et~al.(2007){Domiciano de Souza}, {Driebe},
  {Chesneau} et~al.}]{2007A&A...464...81D}
{Domiciano de Souza}, A., {Driebe}, T., {Chesneau}, O., et~al. 2007, \aap, 464,
  81

\bibitem[{{Dunstall} et~al.(2012){Dunstall}, {Fraser}, {Clark}
  et~al.}]{2012A&A...542A..50D}
{Dunstall}, P.~R., {Fraser}, M., {Clark}, J.~S., et~al. 2012, \aap, 542, A50

\bibitem[{{Ekstr{\"o}m} et~al.(2012){Ekstr{\"o}m}, {Georgy}, {Eggenberger}
  et~al.}]{2012A&A...537A.146E}
{Ekstr{\"o}m}, S., {Georgy}, C., {Eggenberger}, P., et~al. 2012, \aap, 537,
  A146

\bibitem[{{Geisel}(1970)}]{1970ApJ...161L.105G}
{Geisel}, S.~L. 1970, \apjl, 161, L105

\bibitem[{{Graus} et~al.(2012){Graus}, {Lamb}, \& {Oey}}]{2012ApJ...759...10G}
{Graus}, A.~S., {Lamb}, J.~B., \& {Oey}, M.~S. 2012, \apj, 759, 10

\bibitem[{{Gummersbach} et~al.(1995){Gummersbach}, {Zickgraf}, \&
  {Wolf}}]{1995A&A...302..409G}
{Gummersbach}, C.~A., {Zickgraf}, F.-J., \& {Wolf}, B. 1995, \aap, 302, 409

\bibitem[{{Kamath} et~al.(2015){Kamath}, {Wood}, \& {Van
  Winckel}}]{2015MNRAS.454.1468K}
{Kamath}, D., {Wood}, P.~R., \& {Van Winckel}, H. 2015, \mnras, 454, 1468

\bibitem[{{Kastner} et~al.(2010){Kastner}, {Buchanan}, {Sahai}, {Forrest}, \&
  {Sargent}}]{Kastner2010}
{Kastner}, J.~H., {Buchanan}, C., {Sahai}, R., {Forrest}, W.~J., \& {Sargent},
  B.~A. 2010, \aj, 139, 1993

\bibitem[{{Kraus}(2006)}]{2006A&A...456..151K}
{Kraus}, M. 2006, \aap, 456, 151

\bibitem[{{Kraus}(2009)}]{2009A&A...494..253K}
--- 2009, \aap, 494, 253

\bibitem[{{Kraus} et~al.(2007){Kraus}, {Borges Fernandes}, \& {de
  Ara{\'u}jo}}]{2007A&A...463..627K}
{Kraus}, M., {Borges Fernandes}, M., \& {de Ara{\'u}jo}, F.~X. 2007, \aap, 463,
  627

\bibitem[{{Kraus} et~al.(2010){Kraus}, {Borges Fernandes}, \& {de
  Ara{\'u}jo}}]{2010A&A...517A..30K}
--- 2010, \aap, 517, A30

\bibitem[{{Kraus} et~al.(2005){Kraus}, {Borges Fernandes}, {de Ara{\'u}jo}, \&
  {Lamers}}]{2005A&A...441..289K}
{Kraus}, M., {Borges Fernandes}, M., {de Ara{\'u}jo}, F.~X., \& {Lamers},
  H.~J.~G.~L.~M. 2005, \aap, 441, 289

\bibitem[{{Kraus} et~al.(2008){Kraus}, {Borges Fernandes}, {Kub{\'a}t}, \& {de
  Ara{\'u}jo}}]{2008A&A...487..697K}
{Kraus}, M., {Borges Fernandes}, M., {Kub{\'a}t}, J., \& {de Ara{\'u}jo}, F.~X.
  2008, \aap, 487, 697

\bibitem[{{Kraus} et~al.(2013{\natexlab{a}}){Kraus}, {Cidale}, {Arias}
  et~al.}]{2013msao.confE.160K}
{Kraus}, M., {Cidale}, L.~S., {Arias}, M.~L., et~al. 2013{\natexlab{a}}, in
  Massive Stars: From alpha to Omega, 160

\bibitem[{{Kraus} et~al.(2014){Kraus}, {Cidale}, {Arias}
  et~al.}]{2014ApJ...780L..10K}
--- 2014, \apjl, 780, L10

\bibitem[{{Kraus} et~al.(2016){Kraus}, {Cidale}, {Arias}
  et~al.}]{2016A&A...593..112K}
--- 2016, \aap, 593, A112

\bibitem[{{Kraus} et~al.(2015{\natexlab{a}}){Kraus}, {Haucke}, {Cidale}
  et~al.}]{2015A&A...581A..75K}
{Kraus}, M., {Haucke}, M., {Cidale}, L.~S., et~al. 2015{\natexlab{a}}, \aap,
  581, A75

\bibitem[{{Kraus} \& {Lamers}(2003)}]{2003A&A...405..165K}
{Kraus}, M., \& {Lamers}, H.~J.~G.~L.~M. 2003, \aap, 405, 165

\bibitem[{{Kraus} et~al.(2015{\natexlab{b}}){Kraus}, {Oksala}, {Cidale}
  et~al.}]{2015ApJ...800L..20K}
{Kraus}, M., {Oksala}, M.~E., {Cidale}, L.~S., et~al. 2015{\natexlab{b}},
  \apjl, 800, L20

\bibitem[{{Kraus} et~al.(2013{\natexlab{b}}){Kraus}, {Oksala}, {Nickeler}
  et~al.}]{2013A&A...549A..28K}
{Kraus}, M., {Oksala}, M.~E., {Nickeler}, D.~H., et~al. 2013{\natexlab{b}},
  \aap, 549, A28

\bibitem[{{Krti{\v c}ka} et~al.(2015){Krti{\v c}ka}, {Kurf{\"u}rst}, \&
  {Krti{\v c}kov{\'a}}}]{2015A&A...573A..20K}
{Krti{\v c}ka}, J., {Kurf{\"u}rst}, P., \& {Krti{\v c}kov{\'a}}, I. 2015, \aap,
  573, A20

\bibitem[{{Kurf{\"u}rst} et~al.(2014){Kurf{\"u}rst}, {Feldmeier}, \& {Krti{\v
  c}ka}}]{2014A&A...569A..23K}
{Kurf{\"u}rst}, P., {Feldmeier}, A., \& {Krti{\v c}ka}, J. 2014, \aap, 569, A23

\bibitem[{{Lamers} et~al.(1998){Lamers}, {Zickgraf}, {de Winter}
  et~al.}]{Lamers1998}
{Lamers}, H.~J.~G.~L.~M., {Zickgraf}, F.-J., {de Winter}, D., et~al. 1998,
  \aap, 340, 117

\bibitem[{{Langer} \& {Heger}(1998)}]{1998ASSL..233..235L}
{Langer}, N., \& {Heger}, A. 1998, in B[e] stars, edited by A.~M. {Hubert}, \&
  C.~{Jaschek}, vol. 233 of Astrophysics and Space Science Library, 235

\bibitem[{{Leedj{\"a}rv} et~al.(2016){Leedj{\"a}rv}, {G{\'a}lis}, {Hric},
  {Merc}, \& {Burmeister}}]{2016MNRAS.456.2558L}
{Leedj{\"a}rv}, L., {G{\'a}lis}, R., {Hric}, L., {Merc}, J., \& {Burmeister},
  M. 2016, \mnras, 456, 2558

\bibitem[{{Levato} et~al.(2014){Levato}, {Miroshnichenko}, \&
  {Saffe}}]{2014A&A...568A..28L}
{Levato}, H., {Miroshnichenko}, A.~S., \& {Saffe}, C. 2014, \aap, 568, A28

\bibitem[{{Liermann} et~al.(2010){Liermann}, {Kraus}, {Schnurr}, \&
  {Fernandes}}]{2010MNRAS.408L...6L}
{Liermann}, A., {Kraus}, M., {Schnurr}, O., \& {Fernandes}, M.~B. 2010, \mnras,
  408, L6

\bibitem[{{Liermann} et~al.(2014){Liermann}, {Schnurr}, {Kraus}
  et~al.}]{2014MNRAS.443..947L}
{Liermann}, A., {Schnurr}, O., {Kraus}, M., et~al. 2014, \mnras, 443, 947

\bibitem[{{Magalhaes}(1992)}]{1992ApJ...398..286M}
{Magalhaes}, A.~M. 1992, \apj, 398, 286

\bibitem[{{Maravelias} et~al.(2015){Maravelias}, {Kraus}, \&
  {Aret}}]{2015EAS....71..229M}
{Maravelias}, G., {Kraus}, M., \& {Aret}, A. 2015, in EAS Publications Series,
  vol.~71 of EAS Publications Series, 229

\bibitem[{{Maravelias} et~al.(2014){Maravelias}, {Zezas}, {Antoniou}, \&
  {Hatzidimitriou}}]{2014MNRAS.438.2005M}
{Maravelias}, G., {Zezas}, A., {Antoniou}, V., \& {Hatzidimitriou}, D. 2014,
  \mnras, 438, 2005

\bibitem[{{Massey} et~al.(2007){Massey}, {McNeill}, {Olsen}
  et~al.}]{2007AJ....134.2474M}
{Massey}, P., {McNeill}, R.~T., {Olsen}, K.~A.~G., et~al. 2007, \aj, 134, 2474

\bibitem[{{McGregor} et~al.(1988{\natexlab{a}}){McGregor}, {Hyland}, \&
  {Hillier}}]{1988ApJ...324.1071M}
{McGregor}, P.~J., {Hyland}, A.~R., \& {Hillier}, D.~J. 1988{\natexlab{a}},
  \apj, 324, 1071

\bibitem[{{McGregor} et~al.(1988{\natexlab{b}}){McGregor}, {Hyland}, \&
  {Hillier}}]{1988ApJ...334..639M}
--- 1988{\natexlab{b}}, \apj, 334, 639

\bibitem[{{McGregor} et~al.(1989){McGregor}, {Hyland}, \&
  {McGinn}}]{1989A&A...223..237M}
{McGregor}, P.~J., {Hyland}, A.~R., \& {McGinn}, M.~T. 1989, \aap, 223, 237

\bibitem[{{Millour} et~al.(2011){Millour}, {Meilland}, {Chesneau}
  et~al.}]{2011A&A...526A.107M}
{Millour}, F., {Meilland}, A., {Chesneau}, O., et~al. 2011, \aap, 526, A107

\bibitem[{{Morris} et~al.(1996){Morris}, {Eenens}, {Hanson}
  et~al.}]{1996ApJ...470..597M}
{Morris}, P.~W., {Eenens}, P.~R.~J., {Hanson}, M.~M., et~al. 1996, \apj, 470,
  597

\bibitem[{{Muratore} et~al.(2012{\natexlab{a}}){Muratore}, {de Wit}, {Kraus}
  et~al.}]{2012ASPC..464...67M}
{Muratore}, M.~F., {de Wit}, W.~J., {Kraus}, M., et~al. 2012{\natexlab{a}}, in
  Circumstellar Dynamics at High Resolution, edited by A.~C. {Carciofi}, \&
  T.~{Rivinius}, vol. 464 of ASP Conf. Ser., 67

\bibitem[{{Muratore} et~al.(2012{\natexlab{b}}){Muratore}, {Kraus}, \& {de
  Wit}}]{2012BAAA...55..123M}
{Muratore}, M.~F., {Kraus}, M., \& {de Wit}, W.~J. 2012{\natexlab{b}}, Boletin
  de la Asociacion Argentina de Astronomia La Plata Argentina, 55, 123

\bibitem[{{Muratore} et~al.(2015){Muratore}, {Kraus}, {Oksala}
  et~al.}]{2015AJ....149...13M}
{Muratore}, M.~F., {Kraus}, M., {Oksala}, M.~E., et~al. 2015, \aj, 149, 13

\bibitem[{{Oksala} et~al.(2012){Oksala}, {Kraus}, {Arias}
  et~al.}]{2012MNRAS.426L..56O}
{Oksala}, M.~E., {Kraus}, M., {Arias}, M.~L., et~al. 2012, \mnras, 426, L56

\bibitem[{{Oksala} et~al.(2013){Oksala}, {Kraus}, {Cidale}
  et~al.}]{2013A&A...558A..17O}
{Oksala}, M.~E., {Kraus}, M., {Cidale}, L.~S., et~al. 2013, \aap, 558, A17

\bibitem[{{Owocki} et~al.(1996){Owocki}, {Cranmer}, \&
  {Gayley}}]{1996ApJ...472L.115O}
{Owocki}, S.~P., {Cranmer}, S.~R., \& {Gayley}, K.~G. 1996, \apjl, 472, L115

\bibitem[{{Pelupessy} et~al.(2000){Pelupessy}, {Lamers}, \&
  {Vink}}]{2000A&A...359..695P}
{Pelupessy}, I., {Lamers}, H.~J.~G.~L.~M., \& {Vink}, J.~S. 2000, \aap, 359,
  695

\bibitem[{{Podsiadlowski} et~al.(2006){Podsiadlowski}, {Morris}, \&
  {Ivanova}}]{2006ASPC..355..259P}
{Podsiadlowski}, P., {Morris}, T.~S., \& {Ivanova}, N. 2006, in Stars with the
  B[e] Phenomenon, edited by M.~{Kraus}, \& A.~S. {Miroshnichenko}, vol. 355 of
  ASP Conf. Ser., 259

\bibitem[{{Porter}(2003)}]{2003A&A...398..631P}
{Porter}, J.~M. 2003, \aap, 398, 631

\bibitem[{{Saio} et~al.(2013){Saio}, {Georgy}, \&
  {Meynet}}]{2013MNRAS.433.1246S}
{Saio}, H., {Georgy}, C., \& {Meynet}, G. 2013, \mnras, 433, 1246

\bibitem[{{Torres} et~al.(2012){Torres}, {Kraus}, {Cidale}
  et~al.}]{2012MNRAS.427L..80T}
{Torres}, A.~F., {Kraus}, M., {Cidale}, L.~S., et~al. 2012, \mnras, 427, L80

\bibitem[{{Wang} et~al.(2012){Wang}, {Weigelt}, {Kreplin}
  et~al.}]{2012A&A...545L..10W}
{Wang}, Y., {Weigelt}, G., {Kreplin}, A., et~al. 2012, \aap, 545, L10

\bibitem[{{Wheelwright} et~al.(2012{\natexlab{a}}){Wheelwright}, {de Wit},
  {Oudmaijer}, \& {Vink}}]{2012A&A...538A...6W}
{Wheelwright}, H.~E., {de Wit}, W.~J., {Oudmaijer}, R.~D., \& {Vink}, J.~S.
  2012{\natexlab{a}}, \aap, 538, A6

\bibitem[{{Wheelwright} et~al.(2012{\natexlab{b}}){Wheelwright}, {de Wit},
  {Weigelt} et~al.}]{2012A&A...543A..77W}
{Wheelwright}, H.~E., {de Wit}, W.~J., {Weigelt}, G., et~al.
  2012{\natexlab{b}}, \aap, 543, A77

\bibitem[{{Wisniewski} et~al.(2007){Wisniewski}, {Bjorkman}, {Bjorkman}, \&
  {Clampin}}]{2007ApJ...670.1331W}
{Wisniewski}, J.~P., {Bjorkman}, K.~S., {Bjorkman}, J.~E., \& {Clampin}, M.
  2007, \apj, 670, 1331

\bibitem[{{Zickgraf}(2000)}]{2000ASPC..214...26Z}
{Zickgraf}, F. 2000, in IAU Colloq. 175: The Be Phenomenon in Early-Type Stars,
  edited by M.~A. {Smith}, H.~F. {Henrichs}, \& J.~{Fabregat}, vol. 214 of ASP
  Conf. Ser., 26

\bibitem[{{Zickgraf}(2006)}]{2006ASPC..355..135Z}
{Zickgraf}, F.-J. 2006, in Stars with the B[e] Phenomenon, edited by
  M.~{Kraus}, \& A.~S. {Miroshnichenko}, vol. 355 of ASP Conf. Ser., 135

\bibitem[{{Zickgraf} \& {Schulte-Ladbeck}(1989)}]{1989A&A...214..274Z}
{Zickgraf}, F.-J., \& {Schulte-Ladbeck}, R.~E. 1989, \aap, 214, 274

\bibitem[{{Zickgraf} et~al.(1986){Zickgraf}, {Wolf}, {Leitherer}
  et~al.}]{1986A&A...163..119Z}
{Zickgraf}, F.-J., {Wolf}, B., {Leitherer}, C., et~al. 1986, \aap, 163, 119

\bibitem[{{Zickgraf} et~al.(1989){Zickgraf}, {Wolf}, {Stahl}, \&
  {Humphreys}}]{1989A&A...220..206Z}
{Zickgraf}, F.-J., {Wolf}, B., {Stahl}, O., \& {Humphreys}, R.~M. 1989, \aap,
  220, 206

\bibitem[{{Zickgraf} et~al.(1985){Zickgraf}, {Wolf}, {Stahl}
  et~al.}]{1985A&A...143..421Z}
{Zickgraf}, F.-J., {Wolf}, B., {Stahl}, O., et~al. 1985, \aap, 143, 421

\bibitem[{{Zsarg{\'o}} et~al.(2008){Zsarg{\'o}}, {Hillier}, \&
  {Georgiev}}]{2008A&A...478..543Z}
{Zsarg{\'o}}, J., {Hillier}, D.~J., \& {Georgiev}, L.~N. 2008, \aap, 478, 543

\end{thebibliography}

\end{document}